\documentclass[aps,prl,twocolumn]{revtex4}
\usepackage{bm,color,amsmath,amssymb,mathrsfs,latexsym,graphicx,psfrag}

\newcommand{\re}{\mathrm{Re}}
\newcommand{\im}{\mathrm{Im}}

\newcommand{\bk}{{\mathbf k}}
\newcommand{\bq}{{\mathbf q}}
\newcommand{\br}{{\mathbf{r}}}
\newcommand{\be}{\begin{equation}}
\newcommand{\ee}{\end{equation}}

\def\be{\begin{equation}}
\def\ee{\end{equation}}
\def\bea{\begin{eqnarray}}
\def\eea{\end{eqnarray}}

\begin{document}
\title{Tri-Dirac Surface Modes in Topological Superconductors}
\author{Chen Fang$^{1,2,3}$, B. Andrei Bernevig$^3$ and Matthew J. Gilbert$^{2,4,5}$}
\date{\today}
\affiliation{$^{1}$Department of Physics, University of Illinois, Urbana IL 61801-3080}
\affiliation{$^2$Micro and Nanotechnology Laboratory, University of Illinois, Urbana IL 61801}
\affiliation{$^3$Department of Physics, Princeton University, Princeton NJ 08544}
\affiliation{$^4$Department of Electrical and Computer Engineering, University of Illinois, Urbana IL 61801}
\affiliation{$^5$Department of Electronics Engineering, University of Rome "Tor Vergata", Rome, Italy 00133}
\begin{abstract}
We propose a new type of topological surface modes having cubic dispersion in three-dimensional topological superconductors. Lower order dispersions are prohibited by the threefold rotational symmetry and time-reversal symmetry. Cooper pairing in the bulk changes sign under improper rotations, akin to$^{3}$He-B. The surface manifestations are a divergent surface density of states at the Fermi level and isospins that rotate three times as they circle the origin in momentum space. We propose that Heusler alloys with band inversion are candidate materials to harbor the novel topological superconductivity.
\end{abstract}
\maketitle

Topological superconductors (TSC) are a novel class of superconductors where Fermi surface topology and unconventional Cooper pairing in the bulk lead to gapless surface excitations called Majorana fermions\cite{Bernevig2013} - aptly named\cite{Kitaev2001} as the particle-hole symmetry (PHS) renders their on-site creation and annihilation operators equal. The possibility of finding these new fermions, long-sought in high energy physics, in condensed matter systems has excited a wave of interesting studies\cite{Roy2008,Fu:2008fk,schnyder2008,Kitaev2009,Sato2009,Fu2010,Qi2010,Sau2010,Lutchyn2010,Alicea2011,Mourik2012,Das2012,shivamoggi2013,Fang2013a}.

In three dimensions (3D), TSCs are predicted in doped semiconductors having conduction/valence bands inverted by spin-orbit coupling and Cooper pairing that is odd under inversion and invariant under time-reversal (TRS)\cite{Sato2009,Fu2010,Sasaki2012}. The surface Bogoliubov quasiparticle excitations of these TSCs form linear, spin-split Dirac cones (as opposed to spin degenerate cones seen in graphene)\cite{Roy2008,schnyder2008,Chung2009}. More recently, both theoretical and experimental studies show that crystalline symmetries (with or without TRS) protect new classes of topological phases. In Ref.[\onlinecite{Hsieh2012}], it has been shown that mirror reflection symmetry protects gapless excitations in several IV-VI semiconductors and in Ref.[\onlinecite{Chiu2013,Zhang2013,Ueno2013}], the same symmetry is shown to protect gapless surface modes in superconductors possessing mirror-odd Cooper pairing. It then follows that the only way to open a gap in the system is by spontaneously breaking TRS or by forming a surface topologically ordered state\cite{Vishwanath2013,Fidkowski2013}.

In this Letter, we show that the three-fold rotation symmetry and TRS can protect an exotic type of surface states in TSCs whose dispersion consists of two cubic-dispersing bands touching each other at $\bar\Gamma$ in the surface Brillouin zone (BZ). Lower order dispersions, i.e., linear and quadratic, are excluded by the symmetry group generated by $\{C_3,T\}$, if and only if the doublet at $\bar\Gamma$ of the surface BZ has angular momenta $\pm3/2\hbar$ along the normal direction. The surface density of states (DOS) at the Fermi level diverges due to the cubic dispersion, and, by writing down a generic $k\cdot{p}$-model, we see that the spin polarization makes three full rotations as a wave packet traces an iso-energy contour. We then determine the requirement on the bulk superconductivity for these surface modes to appear. We show that the Fermi level in the bulk must cross bands formed by the $j_z=\pm3/2$ states ($\hbar\equiv1$ hereafter), branching from a $\Gamma_8$-representation (denoting the four $p$-states with total angular momentum $j=3/2$ in a spin-orbit split system) on a cubic lattice. This requirement is met in a series of Heusler alloys that are zero gap semiconductors\cite{Goll2008,Lin:2010,Chadov2010,Butch2011}. Furthermore, the Cooper pairs are required to transform nontrivially under the cubic symmetry group: they change sign under improper rotations but not under proper ones. We perform a concrete model study of a spin-$3/2$ Fermi liquid with full $O(3)$ symmetry and TRS, showing its leading instability towards this nontrivial singlet Cooper pairing, induced by screened Coulomb repulsion. The resultant superconductivity is a close analogue of the superfluity in the B-phase of $^3$He\cite{Leggett1975,Vollhardt1990}, but in a Fermi liquid whose constituent particle is spin-$3/2$. Finally, we discuss experimental signatures that characterize the new topological superconductor.

\begin{figure}
\includegraphics[width=8cm]{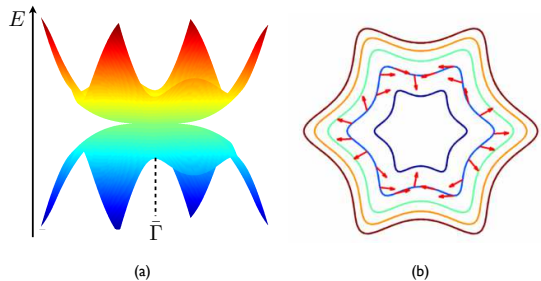}
\caption{(a) The energy dispersion of the Bogoliubov excitations around a tri-Dirac point. (b) The iso-energy contours and the pseudospin structure of $h_{3/2}(\bk)$ along the contour taking $c_2=2c_3=1$.}
\label{fig:dispersion}
\end{figure}

We start by considering a two-band $k\cdot{p}$-theory for the surface states of a TSC, in the Bogoliubov-de Gennes (BdG) form, around $\bar\Gamma$ where a doublet of Majorana modes are \emph{assumed} to exist. The symmetries are the TRS, denoted by $T$, PHS, by $P$, and the three-fold rotation $C_3$. They commute with each other, as they act on different degrees of freedom of time, charge and space, respectively. The spin-orbit coupling has broken the SU(2) symmetry of spin rotation so $C_3$ simultaneously rotates the position and the spin of a particle. For odd-integer spins, $C_3^3=-1$ due to the Berry phase brought by the spin. These constraints result in the following irreducible representations of the group generators (up to an arbitrary unitary transformation):
\bea\label{eq:irrep}
E_{1/2}:\;T&=&K(i\sigma_y),\;P=K\sigma_x,\;C_3=e^{i\sigma_z\pi/3},\\
\nonumber
E_{3/2}:\;T&=&K(i\sigma_y),\;P=K\sigma_x,\;C_3=-I_{2\times2}.
\eea
In Eq.(\ref{eq:irrep}), $C_3$-rotation operator is $\exp(i\hat{j}_z\frac{2\pi}{3})$ with $\hat{j}_z=\sigma_z/2$ ($\hat{j}_z=3\sigma_z/2$) for the $E_{1/2}$ ($E_{3/2}$). The $k\cdot{p}$-Hamiltonian, $h(\bk)$, must satisfy the following symmetry constraints:
\bea\label{eq:constraints}
Th(\bk)T^{-1}&=&h(-\bk),\\
\nonumber Ph(\bk)P^{-1}&=&-h(-\bk),\\
\nonumber C_3h(k_+,k_-)C_3^{-1}&=&h(k_+e^{i2\pi/3},k_-e^{-i2\pi/3}),
\eea
where $k_\pm=k_x\pm{i}k_y$. Using Eqs.(\ref{eq:irrep},\ref{eq:constraints}), we obtain
\bea\label{eq:effective}
&&h_{1/2}(\bk)=\re[c_1k_+]\sigma_x+\im[c_1k_+]\sigma_y,\\\nonumber
&&h_{3/2}(\bk)=\re[c_2k^3_++c_3k^3_-]\sigma_x+\im[c_2k^3_++c_3k^3_-]\sigma_y,
\eea
where $c_{1,2}$ are complex numbers that are material dependent. Hence, we find that while the dispersion of $h_{1/2}$ is linear (Dirac-like) around $\bk=0$, the dispersion of $h_{3/2}$ is cubic, $E_{3/2}(\bk)=|c_2k_+^3+c_3k_-^3|$. Linear and second order terms in $\bk$ do not appear [see Fig.\ref{fig:dispersion}(a)]. This can be understood by examining the continuum limit with $C_\infty$-symmetry: a hopping from a state of $j_z=\pm3/2$ to $j_z=\mp3/2$ changes the total angular momentum by $\mp3$, so in order to preserve the rotation symmetry, this hopping should be ``balanced" by an orbital part of $k_\pm^3$.

The wavefunction of $h_{3/2}(\bk)$ in Eq.(\ref{eq:effective}) is described by the pseudospin structure at each $\bk$. The pseudo-spin up (down) states correspond to the first (second) basis vector in the $k\cdot{p}$-model, i.e., the two degenerate states at $\bk=0$. The pseudo-spin at any $\bk$ is then given by a unit vector in the $xy$-plane, as is guaranteed by PHS, whose two components are $(S_x,S_y)=(\re[c_2k_+^3+c_3k_-^3]/|c_2k_+^3+c_3k_-^3|,\im[c_2k_+^3+c_3k_-^3]/|c_2k_+^3+c_3k_-^3|)$. As the momentum makes a full circle enclosing the origin clockwise, the pseudo-spin has made three full rotations clockwise (if $|c_2|>|c_3|$) or counterclockwise (if $|c_2|<|c_3|$) [see Fig.\ref{fig:dispersion}(b)]. Therefore, we refer to the degeneracy point at $h_{3/2}(\bk)$ in Eq.(\ref{eq:effective}) at $\bk=0$ the \emph{tri-Dirac point}, as the evolution of the wavefunction around is topologically equivalent to that around three Dirac points, resulting in total winding number of $\pm3$. The tri-Dirac point splits if $C_3$ is broken by, for instance, mechanical strain. While a generic band crossing with linear dispersion (Dirac point) has vanishing density of states (DOS), the DOS at a tri-Dirac point is divergent, given by $\rho(E)=\frac{1}{2}\int\delta(E_{3/2}(\bk)-E){d\bk^2}\propto{E}^{-1/3}$, where the prefactor $1/2$ is because each excitation close to the Fermi level is roughly an equal weight linear superposition of electron and hole states.

One may naturally infer from the divergence of DOS at the band touching that the residual interaction between Bogoliubov excitations may be relevant and open a gap in the spectrum. As the surface states are of Majorana character, they contain two species of Bogoliubov excitations, namely $\gamma_{1,2}(\bk)$, satisfying $\gamma_i(\bk)=\gamma^\dag_i(-\bk)$ in $\bk$-space, or,  $\gamma_i(\br)=\gamma^\dag_i(\br)$ as a real field in real space. The interaction hence must contain at least two spatial derivatives, such as $-g\gamma_1(\br)\gamma_2(\br)\nabla\gamma_1(\br)\cdot\nabla\gamma_2(\br)$. Simple dimension counting shows that coupling constant $g$ is irrelevant, meaning that it flows to zero under renormalization towards the long wavelength limit. The surface states of a tri-Dirac cone are hence robust against weak residual interactions between quasiparticles.

With the nature of the surface states understood, we now seek the form of the bulk superconductivity that gives rise to this new type of surface states. We begin by noticing that since the tri-Dirac point derives its protection from both $C_3$-symmetry and TRS, it may only appear on terminations that preserve both these symmetries, i.e. at the $\bar\Gamma$ point, in the surface BZ. (Other $C_3$-invariant points $\bar{K}$ and $\bar{K}'$ which are invariant under $C_2*T$ do not have degeneracy because $(C_2*T)^2=+1$.) Second, we need a pair of Majorana modes at $\bar\Gamma$, which depends on band structure along the line, parameterized by $k_z$, in the 3D BZ that projects onto $\bar\Gamma$. In the weak coupling limit, the topology of this line depends on the signs of pairing amplitude at the Fermi points where the line crosses the Fermi surface: it is nontrivial/trivial if there are odd/even number of Fermi points that have \emph{negative} pairing signs in the region $k_z>0$. Finally, we need the Majorana modes to have angular momenta $\pm3/2$. In the weak coupling limit, the surface Bogoliubov excitations are linear combinations of particle- and hole-states on the Fermi surface in the bulk, so we need the Fermi level to cross bulk bands that have angular momenta $\pm3/2$. In many Heusler alloys, the $(111)$-terminations are $C_3$-symmetric; the Fermi level is at the $\Gamma_8$-representation, branching into two sets of doublet bands having angular momenta $\pm3/2$ (the $E_{3/2}$-branch as conduction band) and $\pm1/2$ (the $E_{1/2}$-branch as valence band) along $\Gamma{L}$ from $\Gamma$, respectively\cite{Chadov2010}. In a thin-film sample, chemical doping may be applied through liquid gating\cite{Cho2011} to tune the chemical potential into the $j_z=\pm3/2$ bands in this system. The remaining question is what interaction induces a Cooper pairing that changes sign between the two bands having $|j_z|=3/2$? As the Fermi level is close to the $\Gamma_8$-representation, it is reasonable to consider a continuum model with TRS and full $O(3)$ symmetry group, of which the point groups of half Heusler alloys ($T_d$) and full Heusler alloys ($O_h$) are subgroups. The surface states obtained in this symmetry-enhanced model would certainly change as one breaks $O(3)$ down to $T_d$ or $O_h$, but the tri-Dirac point on the surface would not be broken, because only $C_3$ and TRS are needed for its protection. The restoration of $O(3)$ simplifies the calculation and makes underlying physics more transparent. The normal state Hamiltonian for a Fermi gas having $O(3)$ and TRS around $\Gamma_8$-representation is given by the following four-band $k\cdot{p}$-model
\bea
H_0(\bk)&=&(\lambda_1+\frac{5}{2}\lambda_2)k^2-2\lambda_2(\bk\cdot\mathbf{S})^2-\mu,
\eea
where $\lambda_{1,2}$ are Luttinger parameters [e.g., $(\lambda_1,\lambda_2)\sim(-2.5,-3.8)2m_e/\hbar^2$ in ScPtBi\cite{Chadov2010}] and $\mathbf{S}=(S_x,S_y,S_z)$ are the spin operators of a spin-$3/2$ fermion, physically realized by spin-orbital splitting of the $p$-orbitals. The isotropic dispersion of $H_0(\bk)$ is given by $\epsilon_{1/2,3/2}(\bk)=(\lambda_1\pm2\lambda_2)k^2-\mu$, where both bands are doubly degenerate. The general form of short-ranged density interaction is $\hat{V}=\sum_\bq{V}(|\bq|)n(\bq)n(-\bq)$, where $n(\bq)$ is the Fourier transform of local density operator and $V(|\bq|)=V_0-V_2q^2+O(q^4)$ is the Fourier transform of the interaction, and further expansion is ignored in the long wavelength limit ($k_F\ll1$). For a screened Coulomb repulsion $V(r)=\frac{e^2}{r}e^{-r/r_0}$, $V_0=e^2r_0^2$ and $V_2=e^2r_0^4$, where $r_0$ is the screening length. Decomposing the interaction into various channels of instability, we have (modulo non-superconducting channels)
\bea\label{eq:SCchannels}
\hat{V}&=&V_0|\sum_{\bk,m,m'}(1-\frac{V_2}{V_0}k^2)b_{mm'}(\bk)|^2\\
\nonumber&-&2V_2\sum_{\bk_1,\bk_2,m,m'}\bk_1\cdot\bk_2b^\dag_{mm'}(\bk_1)b_{mm'}(\bk_2),
\eea
where $b_{mm'}(\bk)=c_{m}(\bk)c_{m'}(-\bk)$ is the electron pair operator and the spin-index $m$ runs in $(-3/2,-1/2,1/2,3/2)$. All pairing channels decompose into squares of irreducible representations of SO(3), following a formal procedure detailed in the Supplementary Materials. Generally, a Cooper pair of zero momentum is determined by the total spin of the two constituent electrons, $\mathbf{S}$, and the angular momentum describing the relative motion between them, $\mathbf{L}$. In the presence of spin-orbital coupling (SOC), the conserved quantities are therefore $S^2$, $L^2$, $J^2$ and $J_z$, where $\mathbf{J}$ is the total angular momentum of the pair, or, $\mathbf{J}=\mathbf{L}+\mathbf{S}$. Therefore, the order parameters are denoted by three quantum numbers $(L,S,J)$, and the ground state is $2J+1$-fold degenerate. In our system where electrons are spin-$3/2$, the total spin is $S=0,1,2,3$; the orbital angular momentum $L=0,1$ because a pairing order parameter for $L\ge2$ is at least quadratic in $k$, coming from decoupling a quartic term in the expansion of $V(q)$.  As a result, the possible pairings are: $(L,S,J)=(0,0,0)$, $(1,1,0)$, $(1,1,1)$, $(0,2,2)$, $(1,1,2)$, $(1,3,2)$, $(1,3,3)$ and $(1,3,4)$. $(L,S,J)=(0,0,0)$ corresponds to the normal $s$-wave pairing and is invariant under all symmetry operations; $(L,S,J)=(1,1,0)$ pairing is analogous to the pairing in the B-phase of $^3$He, invariant under all proper rotations but changes sign under improper ones (parity odd)\cite{Fu2010,qi2011rev}. Without accurate information on the material dependent interaction parameters, it is hard to explore the complete phase diagram with all types of instabilities, but since all pairings with $J>0$ are nodal (assuming that the FS encloses $\Gamma$) and that nodal gaps are in general less energetically favored than the full gaps, we restrict the discussion to the two singlet pairings:
\bea\label{eq:SCsimplified}
\hat{V}\approx{}\frac{|\hat\Delta_{000}|^2}{V_0}-\frac{2|\hat\Delta_{110}|^2}{3V_2},
\eea
in which
\bea\nonumber
&&\frac{\hat\Delta_{000}}{V_0}\equiv\sum_{\bk}(1-\frac{V_0}{V_2}k^2)\langle3/2,3/2;m,m'|3/2,3/2;0,0\rangle{b}_{mm'}(\bk),\\
&&\label{eq:Delta110}\frac{\hat\Delta_{110}}{V_2}\equiv\sum_{\bk,M,S_z,m,m'}k^{(M)}\langle1,1;M,S_z|1,1;0,0\rangle\times\\
\nonumber&&\langle3/2,3/2;m,m'|3/2,3/2;1,S_z\rangle{b}_{mm'}(\bk),
\eea
where $k^{(\pm1)}\equiv\mp\sqrt{3/2}k_\pm$, $k^{(0)}\equiv\sqrt{3}{k}_z$, and $\langle{j_1},j_2;m_1,m_2|j_1,j_2;J,M\rangle$'s are the Clebsch-Gordan coefficients, expanding the eigenstate of total angular momentum $|J,M\rangle$ in the product basis of $|j_1,m_1\rangle\otimes|j_2,m_2\rangle$. The relative sign between the two terms in Eq.(\ref{eq:SCsimplified}) is because the attraction (repulsion) favors the parity even (odd) pairs where the wavefunctions of the two electrons have greater (smaller) overlap in real space. The interaction leads to spontaneous superconducting ordering of $\hat\Delta_{000}$ and $\hat\Delta_{110}$ if $V_0<0$ (e.g., BCS attraction) and $V_2>0$ (e.g., screened Coulomb repulsion), respectively.

A standard mean field calculation (see Supplementary Materials) shows that when $V_0<0$, $T_{e}>T_{o}$, where $T_{e/o}$ are transition temperatures for even/odd parity superconductivity, meaning that as long as $V_0$ is attractive, the conventional $s$-wave pairing is energetically favored. The odd parity pairing is energetically favored if $V_0,V_2>0$, in other words, the interaction is repulsive. (While the $\Delta_{000}$ channel becomes attractive for $k>V_0/V_2$, this is because we did not keep higher order terms in the expansion of $V(|\bq|)$. Generally, repulsive screened Coulomb potential disfavors pairs of zero angular momentum.)

\begin{figure}
\includegraphics[width=8cm]{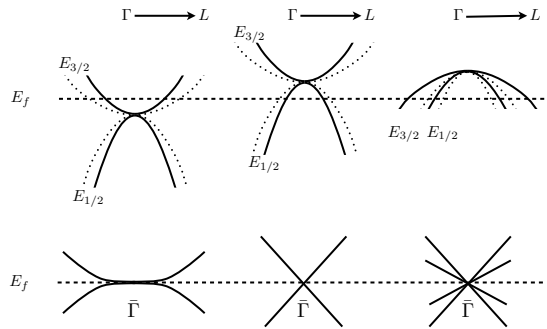}
\caption{Three possible scenarios of the how the Fermi level crosses the bulk $\Gamma_8$-bands and the schematic dispersions of the surface states. Solid and dotted lines mean that the corresponding band of has positive and negative pairing amplitude, respectively, in the nontrivial superconducting phase. Each band here is non-degenerate as inversion symmetry is absent in general. In the presence of inversion symmetry, bands become doubly degenerate; the doublet Fermi surfaces have opposite signs of pairing if the pairing operator changes sign under inversion\cite{Fu2010,Note1}.}
\label{fig:three}
\end{figure}

When the $(L,S,J)=(1,1,0)$ phase is favored by repulsive interaction, there are three possible scenarios (see Fig.\ref{fig:three}), in which the Fermi level crosses: (i) the $E_{1/2}$-bands, (ii) the $E_{3/2}$-bands and (iii) all four $\Gamma_8$ bands. (The $\Gamma_8$ representation also exists in non-centrosymmetric lattices so bands are generally non-degenerate.) In case-(i), there is a Dirac point at the $\bar\Gamma$ with linear dispersion at its vicinity; in case-(ii), the Dirac point is replaced by a tri-Dirac point, having cubic dispersion and divergent DOS; in case-(iii), there are two doublets at $\Gamma$, having angular momentum $j_z=\pm1/2$ and $j_z=\pm3/2$, respectively. Here we note that in case-(iii), all four branches of dispersion are generically linear (see Supplementary Materials for details). It is interesting to consider the $Z$-index of the class DIII superconductors for these cases and one obtains $\pm1$, $\pm3$ and $\pm2$. In the first two cases the Chern numbers of the FS associated with the $E_{1/2,3/2}$-bands are $\pm1$ and $\pm3$. Therefore if the sign of pairing changes between its two pieces, from the Qi-Hughes-Zhang formula\cite{Qi2010} we have $z=\pm1,\pm3$, respectively. For case-(iii), there are a pair of sign-changing $E_{1/2}$-FS and another of sign-changing $E_{3/2}$-FS, so it seems we can have either $\pm2$ or $\pm4$. An expansion of $\hat\Delta_{110}$ in Eq.(\ref{eq:Delta110}) shows a relative minus sign between the pairing on $E_{1/2}$- and $E_{3/2}$-bands, leaving $\pm2$ the only possibility. Therefore, only in case-(ii) does the phase exhibit protected tri-Dirac surface states with cubic dispersion. To realize this scenario, the conduction and the valence bands must touch at $\Gamma$ and bend oppositely, making the normal state a doped zero gap semiconductor.

Heusler alloys offer a wide spectrum of zero gap semiconductors, as a result of the interplay between SOC and orbital hybridization\cite{Chadov2010}. When SOC is stronger than hybridization (which is almost always the case in topological insulators as HgTe), at $\Gamma$ we have the $\Gamma_8$-representation higher than the $\Gamma_6$-representation and the Fermi level of an undoped compound is exactly at the branching point of $E_{3/2}$-bands (conduction) and $E_{1/2}$-bands (valence). When electron-doped, these materials in general have $E_f$ crossing the $E_{3/2}$-bands. The properties of a Heusler alloy strongly depends on its composition, and include: the superconductivity in LaPtBi and YPtBi\cite{Goll2008,Butch2011}, the ferromagnetism in GdPtBi\cite{Canfield1991} and the heavy fermion behavior of YbPtBi\cite{Fisk1991}. Since we adopt the most general form of short range electron interaction in the long wavelength limit (small $|\bq|$), the results suggest that superconducting Heusler alloys such as YPtBi may be candidate materials. In some Heusler alloys where magnetic interactions are significant, spin-spin interaction is not negligible, and the competition between magnetism and superconductivity must be considered.

\begin{figure}
\includegraphics[width=8cm]{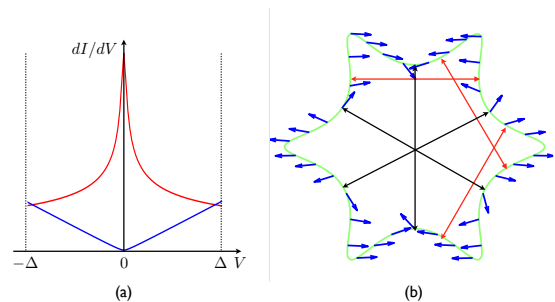}
\caption{(a) Schematic plot of the differential conductance on the surface of TSCs with Dirac (blue) and tri-Dirac surface states inside the bulk superconducting gap. (b) Typical symmetry-forbidden scattering vectors on the iso-energy contour of the tri-Dirac surface states. Black arrows indicate scattering vectors forbidden by TRS, and red arrows are vectors strongly suppressed due to the special pseudospin structure of tri-Dirac surface states, over-plotted on the contour.}
\label{fig:forbidden}
\end{figure}

Experimentally, the new TSCs can be identified by measuring its surface DOS using STM. The $dI/dV$ curve diverges as $dI/dV\propto{}V^{-1/3}$ in the vicinity, as opposed to $dI/dV\propto{V}$ in TSCs with simple Dirac surface states [see Fig.\ref{fig:forbidden}(a)]. Quasiparticle interference on the surface can be used to verify the pseudospin structure\cite{roushan2009,Alpichshev2010}. The iso-energy contour is generically a hexagram even at low energy, similar to the contour in the surface states of Bi$_2$Te$_3$ topological insulator away from the linear Dirac regime. In the latter, the strongest peaks in QPI results from scattering vectors indicated by the red arrows in Fig.\ref{fig:forbidden}(b)\cite{Xue:2009,ZhouX2009,Lee:2009}. These peaks indicate strong interference between states having opposite velocity and not related by TRS. On a hexagram contour, the two momenta of the interfering states make an angle of $\sim60^\circ$. However, for the tri-Dirac surface states, these scattering channels are strongly suppressed, because the pseudospins are exactly opposite to each other between states related by a 60-degree rotation, a signature of the new topological surface states.

\textbf{Acknowledgements}The authors thank E. Fradkin for helpful discussions. CF is supported by ONR-N0014-11-1-0728 for salary and DARPA-N66001-11-1-4110 for travel. MJG acknowledges support from the AFOSR under grant FA9550-10-1-0459, the ONR under grant N0014-11-1-0728, a fellowship from the Center for Advanced Study (CAS) at the University of Illinois. BAB was supported by NSF CAREER DMR-095242, ONR-N00014-11-1-0635, Darpa-N66001-11-1-4110, David and Lucile Packard Foundation, MURI-130-6082, NSF-MSREC-339-6225 and SPAWARSYCEN Pacific 339-6455.

\bibliography{TI}

\begin{thebibliography}{39}
\expandafter\ifx\csname natexlab\endcsname\relax\def\natexlab#1{#1}\fi
\expandafter\ifx\csname bibnamefont\endcsname\relax
  \def\bibnamefont#1{#1}\fi
\expandafter\ifx\csname bibfnamefont\endcsname\relax
  \def\bibfnamefont#1{#1}\fi
\expandafter\ifx\csname citenamefont\endcsname\relax
  \def\citenamefont#1{#1}\fi
\expandafter\ifx\csname url\endcsname\relax
  \def\url#1{\texttt{#1}}\fi
\expandafter\ifx\csname urlprefix\endcsname\relax\def\urlprefix{URL }\fi
\providecommand{\bibinfo}[2]{#2}
\providecommand{\eprint}[2][]{\url{#2}}

\bibitem[{\citenamefont{Bernevig and Hughes}(2013)}]{Bernevig2013}
\bibinfo{author}{\bibfnamefont{B.~A.} \bibnamefont{Bernevig}} \bibnamefont{and}
  \bibinfo{author}{\bibfnamefont{T.~L.} \bibnamefont{Hughes}},
  \emph{\bibinfo{title}{Topological Insulators and Topological
  Superconductors}} (\bibinfo{publisher}{Princeton University Press},
  \bibinfo{year}{2013}).

\bibitem[{\citenamefont{Kitaev}(2001)}]{Kitaev2001}
\bibinfo{author}{\bibfnamefont{A.~Y.} \bibnamefont{Kitaev}},
  \bibinfo{journal}{Physics-Uspekhi} \textbf{\bibinfo{volume}{44}},
  \bibinfo{pages}{131} (\bibinfo{year}{2001}).

\bibitem[{\citenamefont{Roy}(2008)}]{Roy2008}
\bibinfo{author}{\bibfnamefont{R.}~\bibnamefont{Roy}},
  \bibinfo{journal}{arXiv:0803.2868}  (\bibinfo{year}{2008}).

\bibitem[{\citenamefont{Fu and Kane}(2008)}]{Fu:2008fk}
\bibinfo{author}{\bibfnamefont{L.}~\bibnamefont{Fu}} \bibnamefont{and}
  \bibinfo{author}{\bibfnamefont{C.~L.} \bibnamefont{Kane}},
  \bibinfo{journal}{Physical Review Letters} \textbf{\bibinfo{volume}{100}}
  (\bibinfo{year}{2008}).

\bibitem[{\citenamefont{Schnyder et~al.}(2008)\citenamefont{Schnyder, Ryu,
  Furusaki, and Ludwig}}]{schnyder2008}
\bibinfo{author}{\bibfnamefont{A.~P.} \bibnamefont{Schnyder}},
  \bibinfo{author}{\bibfnamefont{S.}~\bibnamefont{Ryu}},
  \bibinfo{author}{\bibfnamefont{A.}~\bibnamefont{Furusaki}}, \bibnamefont{and}
  \bibinfo{author}{\bibfnamefont{A.~W.~W.} \bibnamefont{Ludwig}},
  \bibinfo{journal}{Phys. Rev. B} \textbf{\bibinfo{volume}{78}},
  \bibinfo{pages}{195125} (\bibinfo{year}{2008}).

\bibitem[{\citenamefont{Kitaev}(2009)}]{Kitaev2009}
\bibinfo{author}{\bibfnamefont{A.}~\bibnamefont{Kitaev}}, \bibinfo{journal}{AIP
  Conf. Proc.} \textbf{\bibinfo{volume}{1134}}, \bibinfo{pages}{22}
  (\bibinfo{year}{2009}).

\bibitem[{\citenamefont{Sato et~al.}(2009)\citenamefont{Sato, Takahashi, and
  Fujimoto}}]{Sato2009}
\bibinfo{author}{\bibfnamefont{M.}~\bibnamefont{Sato}},
  \bibinfo{author}{\bibfnamefont{Y.}~\bibnamefont{Takahashi}},
  \bibnamefont{and} \bibinfo{author}{\bibfnamefont{S.}~\bibnamefont{Fujimoto}},
  \bibinfo{journal}{Phys. Rev. Lett.} \textbf{\bibinfo{volume}{103}},
  \bibinfo{pages}{020401} (\bibinfo{year}{2009}).

\bibitem[{\citenamefont{Fu and Berg}(2010)}]{Fu2010}
\bibinfo{author}{\bibfnamefont{L.}~\bibnamefont{Fu}} \bibnamefont{and}
  \bibinfo{author}{\bibfnamefont{E.}~\bibnamefont{Berg}},
  \bibinfo{journal}{Phys. Rev. Lett.} \textbf{\bibinfo{volume}{105}},
  \bibinfo{pages}{097001} (\bibinfo{year}{2010}).

\bibitem[{\citenamefont{Qi et~al.}(2010)\citenamefont{Qi, Hughes, and
  Zhang}}]{Qi2010}
\bibinfo{author}{\bibfnamefont{X.-L.} \bibnamefont{Qi}},
  \bibinfo{author}{\bibfnamefont{T.~L.} \bibnamefont{Hughes}},
  \bibnamefont{and} \bibinfo{author}{\bibfnamefont{S.-C.} \bibnamefont{Zhang}},
  \bibinfo{journal}{Phys. Rev. B} \textbf{\bibinfo{volume}{81}},
  \bibinfo{pages}{134508} (\bibinfo{year}{2010}).

\bibitem[{\citenamefont{Sau et~al.}(2010)\citenamefont{Sau, Lutchyn, Tewari,
  and Das~Sarma}}]{Sau2010}
\bibinfo{author}{\bibfnamefont{J.~D.} \bibnamefont{Sau}},
  \bibinfo{author}{\bibfnamefont{R.~M.} \bibnamefont{Lutchyn}},
  \bibinfo{author}{\bibfnamefont{S.}~\bibnamefont{Tewari}}, \bibnamefont{and}
  \bibinfo{author}{\bibfnamefont{S.}~\bibnamefont{Das~Sarma}},
  \bibinfo{journal}{Phys. Rev. Lett.} \textbf{\bibinfo{volume}{104}},
  \bibinfo{pages}{040502} (\bibinfo{year}{2010}).

\bibitem[{\citenamefont{Lutchyn et~al.}(2010)\citenamefont{Lutchyn, Sau, and
  Das~Sarma}}]{Lutchyn2010}
\bibinfo{author}{\bibfnamefont{R.~M.} \bibnamefont{Lutchyn}},
  \bibinfo{author}{\bibfnamefont{J.~D.} \bibnamefont{Sau}}, \bibnamefont{and}
  \bibinfo{author}{\bibfnamefont{S.}~\bibnamefont{Das~Sarma}},
  \bibinfo{journal}{Phys. Rev. Lett.} \textbf{\bibinfo{volume}{105}},
  \bibinfo{pages}{077001} (\bibinfo{year}{2010}).

\bibitem[{\citenamefont{Alicea et~al.}(2011)\citenamefont{Alicea, Oreg, Refael,
  von Oppen, and Fisher}}]{Alicea2011}
\bibinfo{author}{\bibfnamefont{J.}~\bibnamefont{Alicea}},
  \bibinfo{author}{\bibfnamefont{Y.}~\bibnamefont{Oreg}},
  \bibinfo{author}{\bibfnamefont{G.}~\bibnamefont{Refael}},
  \bibinfo{author}{\bibfnamefont{F.}~\bibnamefont{von Oppen}},
  \bibnamefont{and} \bibinfo{author}{\bibfnamefont{M.~P.~A.}
  \bibnamefont{Fisher}}, \bibinfo{journal}{Nat Phys}
  \textbf{\bibinfo{volume}{7}}, \bibinfo{pages}{412} (\bibinfo{year}{2011}).

\bibitem[{\citenamefont{Mourik et~al.}(2012)\citenamefont{Mourik, Zuo, Frolov,
  Plissard, Bakkers, and Kouwenhoven}}]{Mourik2012}
\bibinfo{author}{\bibfnamefont{V.}~\bibnamefont{Mourik}},
  \bibinfo{author}{\bibfnamefont{K.}~\bibnamefont{Zuo}},
  \bibinfo{author}{\bibfnamefont{S.~M.} \bibnamefont{Frolov}},
  \bibinfo{author}{\bibfnamefont{S.~R.} \bibnamefont{Plissard}},
  \bibinfo{author}{\bibfnamefont{E.~P. A.~M.} \bibnamefont{Bakkers}},
  \bibnamefont{and} \bibinfo{author}{\bibfnamefont{L.~P.}
  \bibnamefont{Kouwenhoven}}, \bibinfo{journal}{Science}
  \textbf{\bibinfo{volume}{336}}, \bibinfo{pages}{1003} (\bibinfo{year}{2012}).

\bibitem[{\citenamefont{Das et~al.}(2012)\citenamefont{Das, Ronen, Most, Oreg,
  Heiblum, and Shtrikman}}]{Das2012}
\bibinfo{author}{\bibfnamefont{A.}~\bibnamefont{Das}},
  \bibinfo{author}{\bibfnamefont{Y.}~\bibnamefont{Ronen}},
  \bibinfo{author}{\bibfnamefont{Y.}~\bibnamefont{Most}},
  \bibinfo{author}{\bibfnamefont{Y.}~\bibnamefont{Oreg}},
  \bibinfo{author}{\bibfnamefont{M.}~\bibnamefont{Heiblum}}, \bibnamefont{and}
  \bibinfo{author}{\bibfnamefont{H.}~\bibnamefont{Shtrikman}},
  \bibinfo{journal}{Nat Phys} \textbf{\bibinfo{volume}{8}},
  \bibinfo{pages}{887} (\bibinfo{year}{2012}).

\bibitem[{\citenamefont{Shivamoggi and Gilbert}(2013)}]{shivamoggi2013}
\bibinfo{author}{\bibfnamefont{V.}~\bibnamefont{Shivamoggi}} \bibnamefont{and}
  \bibinfo{author}{\bibfnamefont{M.~J.} \bibnamefont{Gilbert}},
  \bibinfo{journal}{Phys. Rev. B} \textbf{\bibinfo{volume}{88}},
  \bibinfo{pages}{134504} (\bibinfo{year}{2013}).

\bibitem[{\citenamefont{Fang et~al.}(2013)\citenamefont{Fang, Gilbert, and
  Bernevig}}]{Fang2013a}
\bibinfo{author}{\bibfnamefont{C.}~\bibnamefont{Fang}},
  \bibinfo{author}{\bibfnamefont{M.~J.} \bibnamefont{Gilbert}},
  \bibnamefont{and} \bibinfo{author}{\bibfnamefont{B.~A.}
  \bibnamefont{Bernevig}}, \bibinfo{journal}{arXiv:1308.2424}
  (\bibinfo{year}{2013}).

\bibitem[{\citenamefont{Sasaki et~al.}(2012)\citenamefont{Sasaki, Ren, Taskin,
  Segawa, Fu, and Ando}}]{Sasaki2012}
\bibinfo{author}{\bibfnamefont{S.}~\bibnamefont{Sasaki}},
  \bibinfo{author}{\bibfnamefont{Z.}~\bibnamefont{Ren}},
  \bibinfo{author}{\bibfnamefont{A.~A.} \bibnamefont{Taskin}},
  \bibinfo{author}{\bibfnamefont{K.}~\bibnamefont{Segawa}},
  \bibinfo{author}{\bibfnamefont{L.}~\bibnamefont{Fu}}, \bibnamefont{and}
  \bibinfo{author}{\bibfnamefont{Y.}~\bibnamefont{Ando}},
  \bibinfo{journal}{Phys. Rev. Lett.} \textbf{\bibinfo{volume}{109}},
  \bibinfo{pages}{217004} (\bibinfo{year}{2012}).

\bibitem[{\citenamefont{Chung and Zhang}(2009)}]{Chung2009}
\bibinfo{author}{\bibfnamefont{S.~B.} \bibnamefont{Chung}} \bibnamefont{and}
  \bibinfo{author}{\bibfnamefont{S.-C.} \bibnamefont{Zhang}},
  \bibinfo{journal}{Phys. Rev. Lett.} \textbf{\bibinfo{volume}{103}},
  \bibinfo{pages}{235301} (\bibinfo{year}{2009}).

\bibitem[{\citenamefont{Hsieh et~al.}(2012)\citenamefont{Hsieh, Lin, Liu, Duan,
  Bansil, and Fu}}]{Hsieh2012}
\bibinfo{author}{\bibfnamefont{T.}~\bibnamefont{Hsieh}},
  \bibinfo{author}{\bibfnamefont{H.}~\bibnamefont{Lin}},
  \bibinfo{author}{\bibfnamefont{J.}~\bibnamefont{Liu}},
  \bibinfo{author}{\bibfnamefont{W.}~\bibnamefont{Duan}},
  \bibinfo{author}{\bibfnamefont{A.}~\bibnamefont{Bansil}}, \bibnamefont{and}
  \bibinfo{author}{\bibfnamefont{L.}~\bibnamefont{Fu}},
  \bibinfo{journal}{Nature Communications} \textbf{\bibinfo{volume}{3}},
  \bibinfo{pages}{982} (\bibinfo{year}{2012}).

\bibitem[{\citenamefont{Chiu et~al.}(2013)\citenamefont{Chiu, Yao, and
  Ryu}}]{Chiu2013}
\bibinfo{author}{\bibfnamefont{C.-K.} \bibnamefont{Chiu}},
  \bibinfo{author}{\bibfnamefont{H.}~\bibnamefont{Yao}}, \bibnamefont{and}
  \bibinfo{author}{\bibfnamefont{S.}~\bibnamefont{Ryu}},
  \bibinfo{journal}{Phys. Rev. B} \textbf{\bibinfo{volume}{88}},
  \bibinfo{pages}{075142} (\bibinfo{year}{2013}).

\bibitem[{\citenamefont{Zhang et~al.}(2013)\citenamefont{Zhang, Kane, and
  Mele}}]{Zhang2013}
\bibinfo{author}{\bibfnamefont{F.}~\bibnamefont{Zhang}},
  \bibinfo{author}{\bibfnamefont{C.~L.} \bibnamefont{Kane}}, \bibnamefont{and}
  \bibinfo{author}{\bibfnamefont{E.~J.} \bibnamefont{Mele}},
  \bibinfo{journal}{Phys. Rev. Lett.} \textbf{\bibinfo{volume}{111}},
  \bibinfo{pages}{056403} (\bibinfo{year}{2013}).

\bibitem[{\citenamefont{Ueno et~al.}(2013)\citenamefont{Ueno, Yamakage, Tanaka,
  and Sato}}]{Ueno2013}
\bibinfo{author}{\bibfnamefont{Y.}~\bibnamefont{Ueno}},
  \bibinfo{author}{\bibfnamefont{A.}~\bibnamefont{Yamakage}},
  \bibinfo{author}{\bibfnamefont{Y.}~\bibnamefont{Tanaka}}, \bibnamefont{and}
  \bibinfo{author}{\bibfnamefont{M.}~\bibnamefont{Sato}},
  \bibinfo{journal}{Phys. Rev. Lett.} \textbf{\bibinfo{volume}{111}},
  \bibinfo{pages}{087002} (\bibinfo{year}{2013}).

\bibitem[{\citenamefont{Vishwanath and Senthil}(2013)}]{Vishwanath2013}
\bibinfo{author}{\bibfnamefont{A.}~\bibnamefont{Vishwanath}} \bibnamefont{and}
  \bibinfo{author}{\bibfnamefont{T.}~\bibnamefont{Senthil}},
  \bibinfo{journal}{Phys. Rev. X} \textbf{\bibinfo{volume}{3}},
  \bibinfo{pages}{011016} (\bibinfo{year}{2013}).

\bibitem[{\citenamefont{Fidkowski et~al.}(2013)\citenamefont{Fidkowski, Chen,
  and Vishwanath}}]{Fidkowski2013}
\bibinfo{author}{\bibfnamefont{L.}~\bibnamefont{Fidkowski}},
  \bibinfo{author}{\bibfnamefont{X.}~\bibnamefont{Chen}}, \bibnamefont{and}
  \bibinfo{author}{\bibfnamefont{A.}~\bibnamefont{Vishwanath}},
  \bibinfo{journal}{arXiv:1305.5851}  (\bibinfo{year}{2013}).

\bibitem[{\citenamefont{Goll et~al.}(2008)\citenamefont{Goll, Marz, Hamann,
  Tomanic, Grube, Yoshino, and Takabatake}}]{Goll2008}
\bibinfo{author}{\bibfnamefont{G.}~\bibnamefont{Goll}},
  \bibinfo{author}{\bibfnamefont{M.}~\bibnamefont{Marz}},
  \bibinfo{author}{\bibfnamefont{A.}~\bibnamefont{Hamann}},
  \bibinfo{author}{\bibfnamefont{T.}~\bibnamefont{Tomanic}},
  \bibinfo{author}{\bibfnamefont{K.}~\bibnamefont{Grube}},
  \bibinfo{author}{\bibfnamefont{T.}~\bibnamefont{Yoshino}}, \bibnamefont{and}
  \bibinfo{author}{\bibfnamefont{T.}~\bibnamefont{Takabatake}},
  \bibinfo{journal}{Physica B: Condensed Matter}
  \textbf{\bibinfo{volume}{403}}, \bibinfo{pages}{1065 }
  (\bibinfo{year}{2008}), ISSN \bibinfo{issn}{0921-4526},
  \bibinfo{note}{<xocs:full-name>Proceedings of the International Conference on
  Strongly Correlated Electron Systems</xocs:full-name>}.

\bibitem[{\citenamefont{Lin et~al.}(2010)\citenamefont{Lin, Wray, Xia, Xu,
  Cava, Bansil, and Hasan}}]{Lin:2010}
\bibinfo{author}{\bibfnamefont{H.}~\bibnamefont{Lin}},
  \bibinfo{author}{\bibfnamefont{A.}~\bibnamefont{Wray}},
  \bibinfo{author}{\bibfnamefont{Y.}~\bibnamefont{Xia}},
  \bibinfo{author}{\bibfnamefont{S.}~\bibnamefont{Xu}},
  \bibinfo{author}{\bibfnamefont{R.~J.} \bibnamefont{Cava}},
  \bibinfo{author}{\bibfnamefont{A.}~\bibnamefont{Bansil}}, \bibnamefont{and}
  \bibinfo{author}{\bibfnamefont{M.~Z.} \bibnamefont{Hasan}},
  \bibinfo{journal}{Nature Materials} \textbf{\bibinfo{volume}{9}},
  \bibinfo{pages}{546} (\bibinfo{year}{2010}).

\bibitem[{\citenamefont{Chadov et~al.}(2010)\citenamefont{Chadov, Qi, Kubler,
  Fecher, Felser, and Zhang}}]{Chadov2010}
\bibinfo{author}{\bibfnamefont{S.}~\bibnamefont{Chadov}},
  \bibinfo{author}{\bibfnamefont{X.}~\bibnamefont{Qi}},
  \bibinfo{author}{\bibfnamefont{J.}~\bibnamefont{Kubler}},
  \bibinfo{author}{\bibfnamefont{G.~H.} \bibnamefont{Fecher}},
  \bibinfo{author}{\bibfnamefont{C.}~\bibnamefont{Felser}}, \bibnamefont{and}
  \bibinfo{author}{\bibfnamefont{S.-C.} \bibnamefont{Zhang}},
  \bibinfo{journal}{Nature Materials} \textbf{\bibinfo{volume}{9}},
  \bibinfo{pages}{541} (\bibinfo{year}{2010}).

\bibitem[{\citenamefont{Butch et~al.}(2011)\citenamefont{Butch, Syers,
  Kirshenbaum, Hope, and Paglione}}]{Butch2011}
\bibinfo{author}{\bibfnamefont{N.~P.} \bibnamefont{Butch}},
  \bibinfo{author}{\bibfnamefont{P.}~\bibnamefont{Syers}},
  \bibinfo{author}{\bibfnamefont{K.}~\bibnamefont{Kirshenbaum}},
  \bibinfo{author}{\bibfnamefont{A.~P.} \bibnamefont{Hope}}, \bibnamefont{and}
  \bibinfo{author}{\bibfnamefont{J.}~\bibnamefont{Paglione}},
  \bibinfo{journal}{Phys. Rev. B} \textbf{\bibinfo{volume}{84}},
  \bibinfo{pages}{220504} (\bibinfo{year}{2011}).

\bibitem[{\citenamefont{Leggett}(1975)}]{Leggett1975}
\bibinfo{author}{\bibfnamefont{A.~J.} \bibnamefont{Leggett}},
  \bibinfo{journal}{Rev. Mod. Phys.} \textbf{\bibinfo{volume}{47}},
  \bibinfo{pages}{331} (\bibinfo{year}{1975}).

\bibitem[{\citenamefont{Vollhardt and Wolffe}(1990)}]{Vollhardt1990}
\bibinfo{author}{\bibfnamefont{D.}~\bibnamefont{Vollhardt}} \bibnamefont{and}
  \bibinfo{author}{\bibfnamefont{P.}~\bibnamefont{Wolffe}},
  \emph{\bibinfo{title}{The superfluid phases of helium 3}}
  (\bibinfo{publisher}{Taylor \& Francis}, \bibinfo{year}{1990}).

\bibitem[{\citenamefont{Cho et~al.}(2011)\citenamefont{Cho, Butch, Paglione,
  and Fuhrer}}]{Cho2011}
\bibinfo{author}{\bibfnamefont{S.}~\bibnamefont{Cho}},
  \bibinfo{author}{\bibfnamefont{N.~P.} \bibnamefont{Butch}},
  \bibinfo{author}{\bibfnamefont{J.}~\bibnamefont{Paglione}}, \bibnamefont{and}
  \bibinfo{author}{\bibfnamefont{M.~S.} \bibnamefont{Fuhrer}},
  \bibinfo{journal}{Nano Letters} \textbf{\bibinfo{volume}{11}},
  \bibinfo{pages}{1925} (\bibinfo{year}{2011}).

\bibitem[{\citenamefont{Qi and Zhang}(2011)}]{qi2011rev}
\bibinfo{author}{\bibfnamefont{X.~L.} \bibnamefont{Qi}} \bibnamefont{and}
  \bibinfo{author}{\bibfnamefont{S.~C.} \bibnamefont{Zhang}},
  \bibinfo{journal}{Rev. Mod. Phys.} \textbf{\bibinfo{volume}{83}},
  \bibinfo{pages}{1057} (\bibinfo{year}{2011}).

\bibitem[{\citenamefont{Canfield et~al.}(1991)\citenamefont{Canfield, Thompson,
  Beyermann, Lacerda, Hundley, Peterson, Fisk, and Ott}}]{Canfield1991}
\bibinfo{author}{\bibfnamefont{P.~C.} \bibnamefont{Canfield}},
  \bibinfo{author}{\bibfnamefont{J.~D.} \bibnamefont{Thompson}},
  \bibinfo{author}{\bibfnamefont{W.~P.} \bibnamefont{Beyermann}},
  \bibinfo{author}{\bibfnamefont{A.}~\bibnamefont{Lacerda}},
  \bibinfo{author}{\bibfnamefont{M.~F.} \bibnamefont{Hundley}},
  \bibinfo{author}{\bibfnamefont{E.}~\bibnamefont{Peterson}},
  \bibinfo{author}{\bibfnamefont{Z.}~\bibnamefont{Fisk}}, \bibnamefont{and}
  \bibinfo{author}{\bibfnamefont{H.~R.} \bibnamefont{Ott}},
  \bibinfo{journal}{J. Appl. Phys.} \textbf{\bibinfo{volume}{70}},
  \bibinfo{pages}{5800} (\bibinfo{year}{1991}).

\bibitem[{\citenamefont{Fisk et~al.}(1991)\citenamefont{Fisk, Canfield,
  Beyermann, Thompson, Hundley, Ott, Felder, Maple, Lopez de~la Torre, Visani
  et~al.}}]{Fisk1991}
\bibinfo{author}{\bibfnamefont{Z.}~\bibnamefont{Fisk}},
  \bibinfo{author}{\bibfnamefont{P.~C.} \bibnamefont{Canfield}},
  \bibinfo{author}{\bibfnamefont{W.~P.} \bibnamefont{Beyermann}},
  \bibinfo{author}{\bibfnamefont{J.~D.} \bibnamefont{Thompson}},
  \bibinfo{author}{\bibfnamefont{M.~F.} \bibnamefont{Hundley}},
  \bibinfo{author}{\bibfnamefont{H.~R.} \bibnamefont{Ott}},
  \bibinfo{author}{\bibfnamefont{E.}~\bibnamefont{Felder}},
  \bibinfo{author}{\bibfnamefont{M.~B.} \bibnamefont{Maple}},
  \bibinfo{author}{\bibfnamefont{M.~A.} \bibnamefont{Lopez de~la Torre}},
  \bibinfo{author}{\bibfnamefont{P.}~\bibnamefont{Visani}},
  \bibnamefont{et~al.}, \bibinfo{journal}{Phys. Rev. Lett.}
  \textbf{\bibinfo{volume}{67}}, \bibinfo{pages}{3310} (\bibinfo{year}{1991}).

\bibitem[{\citenamefont{Roushan et~al.}(2009)\citenamefont{Roushan, Seo,
  Parker, Hor, Hsieh, Qian, Richardella, Hasan, Cava, and
  Yazdani}}]{roushan2009}
\bibinfo{author}{\bibfnamefont{P.}~\bibnamefont{Roushan}},
  \bibinfo{author}{\bibfnamefont{J.}~\bibnamefont{Seo}},
  \bibinfo{author}{\bibfnamefont{C.~V.} \bibnamefont{Parker}},
  \bibinfo{author}{\bibfnamefont{Y.~S.} \bibnamefont{Hor}},
  \bibinfo{author}{\bibfnamefont{D.}~\bibnamefont{Hsieh}},
  \bibinfo{author}{\bibfnamefont{D.}~\bibnamefont{Qian}},
  \bibinfo{author}{\bibfnamefont{A.}~\bibnamefont{Richardella}},
  \bibinfo{author}{\bibfnamefont{M.~Z.} \bibnamefont{Hasan}},
  \bibinfo{author}{\bibfnamefont{R.~J.} \bibnamefont{Cava}}, \bibnamefont{and}
  \bibinfo{author}{\bibfnamefont{A.}~\bibnamefont{Yazdani}},
  \bibinfo{journal}{Nature} \textbf{\bibinfo{volume}{460}},
  \bibinfo{pages}{1106} (\bibinfo{year}{2009}).

\bibitem[{\citenamefont{Alpichshev et~al.}(2010)\citenamefont{Alpichshev,
  Analytis, Chu, Fisher, Chen, Shen, Fang, and Kapitulnik}}]{Alpichshev2010}
\bibinfo{author}{\bibfnamefont{Z.}~\bibnamefont{Alpichshev}},
  \bibinfo{author}{\bibfnamefont{J.~G.} \bibnamefont{Analytis}},
  \bibinfo{author}{\bibfnamefont{J.-H.} \bibnamefont{Chu}},
  \bibinfo{author}{\bibfnamefont{I.~R.} \bibnamefont{Fisher}},
  \bibinfo{author}{\bibfnamefont{Y.~L.} \bibnamefont{Chen}},
  \bibinfo{author}{\bibfnamefont{Z.~X.} \bibnamefont{Shen}},
  \bibinfo{author}{\bibfnamefont{A.}~\bibnamefont{Fang}}, \bibnamefont{and}
  \bibinfo{author}{\bibfnamefont{A.}~\bibnamefont{Kapitulnik}},
  \bibinfo{journal}{Phys. Rev. Lett.} \textbf{\bibinfo{volume}{104}},
  \bibinfo{pages}{016401} (\bibinfo{year}{2010}).

\bibitem[{\citenamefont{Zhang et~al.}(2009)\citenamefont{Zhang, Cheng, Chen,
  Jia, Ma, He, Wang, Zhang, Dai, Fang et~al.}}]{Xue:2009}
\bibinfo{author}{\bibfnamefont{T.}~\bibnamefont{Zhang}},
  \bibinfo{author}{\bibfnamefont{P.}~\bibnamefont{Cheng}},
  \bibinfo{author}{\bibfnamefont{X.}~\bibnamefont{Chen}},
  \bibinfo{author}{\bibfnamefont{J.-F.} \bibnamefont{Jia}},
  \bibinfo{author}{\bibfnamefont{X.}~\bibnamefont{Ma}},
  \bibinfo{author}{\bibfnamefont{K.}~\bibnamefont{He}},
  \bibinfo{author}{\bibfnamefont{L.}~\bibnamefont{Wang}},
  \bibinfo{author}{\bibfnamefont{H.}~\bibnamefont{Zhang}},
  \bibinfo{author}{\bibfnamefont{X.}~\bibnamefont{Dai}},
  \bibinfo{author}{\bibfnamefont{Z.}~\bibnamefont{Fang}}, \bibnamefont{et~al.},
  \bibinfo{journal}{Phys. Rev. Lett.} \textbf{\bibinfo{volume}{103}},
  \bibinfo{pages}{266803} (\bibinfo{year}{2009}).

\bibitem[{\citenamefont{Zhou et~al.}(2009)\citenamefont{Zhou, Fang, Tsai, and
  Hu}}]{ZhouX2009}
\bibinfo{author}{\bibfnamefont{X.}~\bibnamefont{Zhou}},
  \bibinfo{author}{\bibfnamefont{C.}~\bibnamefont{Fang}},
  \bibinfo{author}{\bibfnamefont{W.-F.} \bibnamefont{Tsai}}, \bibnamefont{and}
  \bibinfo{author}{\bibfnamefont{J.-P.} \bibnamefont{Hu}},
  \bibinfo{journal}{Phys. Rev. B} \textbf{\bibinfo{volume}{80}},
  \bibinfo{pages}{245317} (\bibinfo{year}{2009}).

\bibitem[{\citenamefont{Lee et~al.}(2009)\citenamefont{Lee, Wu, Arovas, and
  Zhang}}]{Lee:2009}
\bibinfo{author}{\bibfnamefont{W.-C.} \bibnamefont{Lee}},
  \bibinfo{author}{\bibfnamefont{C.}~\bibnamefont{Wu}},
  \bibinfo{author}{\bibfnamefont{D.~P.} \bibnamefont{Arovas}},
  \bibnamefont{and} \bibinfo{author}{\bibfnamefont{S.-C.} \bibnamefont{Zhang}},
  \bibinfo{journal}{Phys. Rev. B} \textbf{\bibinfo{volume}{80}},
  \bibinfo{pages}{245439} (\bibinfo{year}{2009}).

\bibitem{Note1} Each band is doubly degenerate. At $\bk$ on the Fermi surface, we pick one eigenstate $|\psi_1(\bk)\rangle$, and the other state is chosen to be $|\psi_2(\bk)\rangle\equiv\hat{P}*\hat{T}|\psi(\bk)\rangle$, where $\hat{P}$ is the inversion operator. The gauge at $-\bk$ is fixed to be $|\psi_{1,2}(-\bk)\rangle\equiv\hat{T}|\psi_{1,2}(\bk)\rangle$. Therefore the generic pairing is expressed by a two-by-two matrix $D_{ij}$. Time-reversal requires that $D$ is a linear combination of $\sigma_{0,x,y,z}$ matrices with real coefficients. If $D\propto\sigma_0$, the pairing operator is invariant under inversion; and if $Tr(D)=0$, the pairing operator changes sign under inversion. The latter case is topologically nontrivial. One can redefine $|\psi_{1,2}(\bk)\rangle$ such that $D$ is diagonal, in other words, either $\sigma_0$ or $\sigma_z$, meaning the two states have same or opposite pairings. This is the accurate meaning of `two Fermi surfaces having same/opposite pairing' in the presence of inversion.

\end{thebibliography}

\onecolumngrid
\begin{appendix}
\section{Decomposition of an electron pair into the basis of irreducible representations of $SO(3)$}

A general pairing operator of zero total momentum is represented by
\bea
\hat{\Delta}^{mm'}=\sum_{\bk}f(\bk)c_m(\bk)c_{m'}(-\bk).
\eea
In a system with symmetry, it is necessary to decompose $\hat\Delta_{mm'}$ into irreducible representations the symmetry group. For $SO(3)$, the procedure goes as follows:

(i) The orbital part, $f(\bk)$, can be expanded in terms of irreducible representations in the function space:
\bea
f(\bk)=\sum_{L,M}f_{LM}(k^2)Y^M_L(\hat\bk),
\eea
where $Y_L^M(\hat\bk)$'s are spherical harmonics and $f_{LM}(k^2)$ is a complex function depending only on the radius of $\bk$.

(ii) The spin part $b_{mm'}(\bk)=c_m(\bk)c_{m'}(-\bk)$ decomposes into irreducible representations in the spin space:
\bea\label{eq:temp1}
b_{mm'}(\bk)=\sum_{S,S_z}\langle{s,s;S,S_z}|s,s;m,m'\rangle{\hat\delta}^{S_z}_{ssS}(\bk),
\eea
where $\hat{\delta}^{S_z}_{ssS}$ is a pair operator annihilating a pair of electrons of total spin $S$ and total spin along $z$-axis $S_z$, at $\pm\bk$. $\hat\delta_{ssS}^{S_z}(\bk)$ is even/odd in $\bk$ if and only if $S=$even/odd, due to Fermi statistics. (In our case, there is $s=3/2$, $m,m'=-3/2,-1/2,1/2,3/2$, $S=0,1,2,3$ and $S_z=-S,-S+1,...,S$.)

(iii) Then we decompose the product of two irreducible representations in the orbital part and the spin part into irreducible representations of fixed total angular momentum and total angular momentum along $z$-axis:
\bea\label{eq:temp2}
\sum_{\hat\bk}Y_L^M(\hat\bk)\hat\delta^{S_z}_{ssS}(\bk)=\sum_{J=|L-S|,...,L+S,J_z=-J,...,+J}\langle{L,S;J,J_z}|L,S;M,S_z\rangle\delta(J_z-M-S_z)\hat{\tilde\Delta}_{LSJ}^{J_z}(k^2),
\eea
where,
\bea\label{eq:temp3}
\hat{\tilde\Delta}_{LSJ}^{J_z}(k^2)\equiv\sum_{M',S'_z}\sum_{\hat\bk}\langle{L,S;M',S'_z}|{L,S;J,J_z}\rangle{Y}_L^{M'}(\hat\bk)\hat\delta^{S'_z}_{ssS}(\bk)
\eea
is the Cooper pair operator with total angular momentum $J$ and total angular momentum along $z$-axis $J_z$. The $\delta$-function in Eq.(\ref{eq:temp2}) is written down explicitly only to show the conservation of total angular momentum - it is already in the definition of the C-G coefficients.

Using Eq.(\ref{eq:temp1},\ref{eq:temp2},\ref{eq:temp3}), we obtain
\bea\label{eq:temp4}
\sum_{\hat\bk}Y^M_L(\hat\bk)b_{mm'}(\bk)=\sum_{S,J,J_z,S_z}\langle{L,S;J,J_z}|L,S;M,S_z\rangle\langle{s,s;S,S_z}|s,s;m,m'\rangle\hat{\tilde{\Delta}}_{LSJ}^{J_z}(k^2).
\eea

Since $Y_L^M(\bk)$ is odd/even in $\bk$ when $L=$odd/even, and $\hat\delta_{ssS}^{S_z}(\bk)$ is odd/even in $\bk$ when $S=$odd/even, $L$ and $S$ must have the same parity, or $\hat{\tilde\Delta}_{LSJ}$ vanishes.

\section{Decomposition of the interaction}

With the technique developed in the previous section, we can now decompose the pairing channels in the interaction into pairings that are irreducible representations of $SO(3)$. Here we only consider an isotropic interaction expanded to the second order in the momentum space as shown in Eq.(\ref{eq:SCsimplified}). 

\subsection{The first term in Eq.(\ref{eq:SCchannels})}

Following the procedure, we rewrite $\sum_{\hat\bk}{b}_{mm'}(\bk)$ as
\bea
&&\sum_{\hat\bk}\sum_{S,S_z}Y_0^0(\hat\bk)\langle{3/2,3/2;S,S_z}|3/2,3/2;m,m'\rangle{\hat\delta}^{S_z}_{\frac{3}{2}\frac{3}{2}S}(\bk)\\
\nonumber
&=&\sum_{\hat\bk}Y_0^0(\hat\bk)[\langle{3/2,3/2;0,0}|3/2,3/2;m,m'\rangle{\hat\delta}^{0}_{\frac{3}{2}\frac{3}{2}0}(\bk)+\sum_{S_z=-2,...,+2}\langle{3/2,3/2;2,S_z}|3/2,3/2;m,m'\rangle{\hat\delta}^{S_z}_{\frac{3}{2}\frac{3}{2}2}(\bk)]\\\nonumber
&=&\langle{3/2,3/2;0,0}|3/2,3/2;m,m'\rangle{\hat{\tilde\Delta}}^{0}_{000}(k^2)+\sum_{J_z=-2,...,+2}\langle{3/2,3/2;2,S_z}|3/2,3/2;m,m'\rangle{\hat{\tilde\Delta}}^{J_z}_{022}(k^2),
\eea
where only $J=even$ terms are included as $J=odd$ terms vanish by Fermi statistics.

The first term in Eq.(\ref{eq:SCchannels}) can thus be put in the form
\bea
\sum_{m,m'}|\sum_\bk(1-\frac{V_2}{V_0}k^2){b}_{mm'}(\bk)|^2=V_0\sum_{S=0,2;J_z=-S,...,S}|\sum_k(1-\frac{V_2}{V_0}k^2)\hat{\tilde\Delta}^{J_z}_{0SS}(k^2)|^2.
\eea
Then after defining
\bea
\hat\Delta_{000}&=&\sum_k(1-\frac{V_2}{V_0}k^2)\hat{\tilde\Delta}^{0}_{000}(k^2)/V_0,\\
\nonumber
\hat\Delta^{J_z}_{022}&=&\sum_k(1-\frac{V_2}{V_0}k^2)\hat{\tilde\Delta}^{J_z}_{022}(k^2)/V_0,
\eea
we have decomposed the first term in Eq.(\ref{eq:SCchannels}) into irreducible representations of $SO(3)$:
\bea
\sum_{mm'}|\sum_\bk{b}_{mm'}(\bk)|^2=(|\hat\Delta_{000}|^2+\sum_{J_z=-2,...,2}|\hat\Delta_{022}^{J_z}|^2)/V_0.
\eea

\subsection{The second term in Eq.(\ref{eq:SCchannels})}

The second term in Eq.(\ref{eq:SCchannels}) can be rewritten as
\bea
-2V_2\sum_{\bk_1,\bk_2,m,m'}\bk_1\cdot\bk_2b_{mm'}(\bk_1)b_{mm'}^\dag(\bk_2)=-\frac{2V_2}{3}\sum_{M,m,m'}|\sum_\bk{k}{Y}_1^{M}(\hat\bk)b_{mm'}(\bk)|^2.
\eea
From Eq.(\ref{eq:temp4}), we have
\bea
\sum_{\hat\bk}Y_1^M(\bk)b_{mm'}(\bk)&=&\sum_{S,S_z,J,J_z,M}\langle1,S;J,J_z|1,S;M,S_z\rangle\langle3/2,3/2;S,S_z|3/2,3/2;m,m'\rangle\hat{\tilde\Delta}_{1SJ}^{J_z}(k^2).
\eea
Using the normalization conditions
\bea
\sum_{m,m'}\langle3/2,3/2;S,S_z|3/2,3/2;m,m'\rangle\langle3/2,3/2;m,m'|3/2,3/2;S',S'_z\rangle&=&\delta_{SS'}\delta_{S_zS'_z},\\
\nonumber
\sum_{M,S_z}\langle1,S;J,J_z|1,S;M,S_z\rangle\langle1,S;M,S_z|1,S;J',J'_z\rangle&=&\delta_{JJ'}\delta_{J_zJ'_z},
\eea
we obtain
\bea
\sum_{M,m,m'}|\sum_\bk{k}_1k_2{Y}_1^{M}(\hat\bk)b_{mm'}(\bk)|^2&=&\sum_{S,J,J_z}\sum_{k_1,k_2}k_1k_2\hat{\tilde\Delta}_{1SJ}^{J_z}(k_1)\hat{\tilde\Delta}_{1SJ}^{J_z\dag}(k_2)\\
\nonumber&=&\sum_{S,J,J_z}|\sum_kk\hat{\tilde\Delta}_{1SJ}^{J_z}(k)|^2.
\eea
Defining
\bea
\Delta_{1SJ}^{J_z}=\sum_kk\hat{\tilde\Delta}_{1SJ}^{J_z}(k)/V_2,
\eea
we have
\bea
-2V_2\sum_{\bk_1,\bk_2,m,m'}\bk_1\cdot\bk_2b_{mm'}(\bk_1)b_{mm'}^\dag(\bk_2)=-\frac{2}{3V_2}\sum_{S=1,3,J=|S-1|,...,|S+1|,J_z}|\Delta_{1SJ}^{J_z}|^2.
\eea

\section{A four-band $k\cdot{p}$-model with TRS and $C_{\infty}$}
In this Appendix we discuss the surface Hamiltonian of case-(iii) discussed in the text (see the right plot of Fig.\ref{fig:three}), where there are four Majorana modes at $\bar\Gamma$, two having $J_z=\pm1/2$ and two having $J_z=\pm3/2$. They are forbidden to hybridize and open a gap at $\bar\Gamma$ due to the rotation symmetry. We choose the basis such that
\bea
j_z&=&\left(\begin{matrix} 
      \frac{1}{2} & 0 & 0 & 0\\
      0 & -\frac{1}{2} & 0 & 0\\
      0 & 0 & \frac{3}{2} & 0\\
      0 & 0 & 0 & -\frac{3}{2}
   \end{matrix}\right),\\
T&=&K\sigma_y\oplus\sigma_y.
\eea
Note that this choice is valid since $\sigma_y\oplus\sigma_y$ is antisymmetric so $T^2=-1$ as required for fermions. Symmetry requires that
\bea
e^{ij_z\theta}H_{eff}(k_+,k_-)e^{-ij_z\theta}&=&H_{eff}(k_+e^{i\theta},k_-e^{-i\theta}),\\
\nonumber
TH_{eff}(\bk)T^{-1}&=&H_{eff}(-\bk).
\eea
To the first order in $\bk$, the Hamiltonian is in the following general form:
\bea
H_{eff}(\bk)=\left(\begin{matrix} 
      0 & c_1k_- & c_2k_+ & 0\\
      c^\ast_1k_+ & 0 & 0 & -c^\ast_2k_-\\
      c^\ast_2k_- & 0 & 0 & 0\\
      0 & -c_2k_+ & 0 & 0
   \end{matrix}\right),
\eea
whose dispersion can be solved as
\bea
E=\pm\sqrt{|c_1|^2+2|c_2|^2\pm|c_1|\sqrt{|c_1|^2+4|c_2|^2}}k/\sqrt{2}.
\eea
All four bands are linear in $\bk$. Therefore, rotation symmetry, or any of its subgroups, does not lead to higher order dispersion in a four-band $k\cdot{p}$ theory with four-fold degenerate Majorana 
states at the origin in two dimensions. $C_n$ is a subgroup of the full $U(1)$ rotation symmetry group, and therefore does not protect higher order band dispersions.

\section{Phase diagram of the superconducting phases}

We use mean-field approximation to obtain the BdG Hamiltonian, and use that to calculate the free energy and hence the phase diagram. The interaction in Eq.(\ref{eq:SCsimplified}) is mean-field decoupled as
\bea
\frac{|\hat\Delta_{000}|^2}{V_0}-\frac{2|\hat\Delta_{110}|^2}{3V_2}&=&\frac{\hat\Delta^\dag_{000}\langle\hat\Delta_{000}\rangle+h.c.}{V_0}-\frac{2(\hat\Delta^\dag_{110}\langle\hat\Delta_{110}\rangle+h.c.)}{3V_2}\\
\nonumber
&+&\frac{2|\langle\hat\Delta_{000}\rangle|^2}{3V_2}-\frac{|\langle\hat\Delta_{110}\rangle|^2}{V_0}\\
\nonumber
&+&\frac{|\hat\Delta_{000}-\langle\hat\Delta_{000}\rangle|^2}{V_0}-\frac{2|\hat\Delta_{110}-\langle\hat\Delta_{110}\rangle|^2}{3V_2}.
\eea
The mean-field decoupled BdG Hamiltonian is obtained by ignoring the last term of fluctuation:
\bea\nonumber
\hat{H}_{BdG}&=&\sum_{\bk\in{BZ}}c^\dag(\bk)H_0(\bk)c(\bk)-\frac{2\Delta_o\hat\Delta^\dag_{110}}{3V_2}+\frac{\Delta_e\hat\Delta^\dag_{000}}{V_0}+h.c.\\
&+&\frac{2|\Delta_o|^2}{3V_2}+\frac{|\Delta_e|^2}{-V_0},
\eea
where we have defined $\Delta_e\equiv\langle\hat{\Delta}_{000}\rangle$ and $\Delta_o\equiv\langle\hat\Delta_{110}\rangle$. The dispersion of the BdG Hamiltonian is very easy to solve by utilizing SO(3) symmetry. Due to the symmetry, the dispersion must be isotropic, and we hence only need to solve it at $\bk=(0,0,k)$. Writing $\hat{H}_{BdG}$ in Nambu basis $\psi(\bk)\equiv(c_{3/2}(\bk),c_{1/2}(\bk),c_{-1/2}(\bk),c_{-3/2}(\bk),c^\dag_{3/2}(-\bk),c^\dag_{1/2}(-\bk),c^\dag_{-1/2}(-\bk),c^\dag_{-3/2}(-\bk))^T$, we have
\bea
H^o_{BdG}(\bk=(0,0,k))&=&\psi^\dag(\bk)   
\left(\begin{matrix} 
      \epsilon_{3/2}(k) & 0 & 0 & 0 & 0 & 0 & 0 & -\frac{\Delta_ok}{\sqrt{5}}\\
      0 & \epsilon_{1/2}(k) & 0 & 0 & 0 & 0 & \frac{\Delta_ok}{\sqrt{45}} & 0\\
      0 & 0 & \epsilon_{1/2}(k) & 0 & 0 & \frac{\Delta_ok}{\sqrt{45}} & 0 & 0\\
      0 & 0 & 0 & \epsilon_{3/2}(k) & -\frac{\Delta_ok}{\sqrt{5}} & 0 & 0 & 0\\
      0 & 0 & 0 & -\frac{\Delta_ok}{\sqrt{5}} & -\epsilon_{3/2}(k) & 0 & 0 & 0\\
      0 & 0 &\frac{\Delta_ok}{\sqrt{45}} & 0 & 0 & -\epsilon_{1/2}(k) & 0 & 0\\
      0 & \frac{\Delta_ok}{\sqrt{45}} & 0 & 0 & 0 & 0 & -\epsilon_{1/2}(k) & 0\\
      -\frac{\Delta_ok}{\sqrt{5}} & 0 & 0 & 0 & 0 & 0 & 0 & -\epsilon_{3/2}(k)
\end{matrix}\right)\psi(\bk),
\eea
\bea
&&\nonumber H^e_{BdG}(\bk=(0,0,k))\\
&=&\nonumber\psi^\dag(\bk)   
\left(\begin{matrix} 
      \epsilon_{3/2}(k) & 0 & 0 & 0 & 0 & 0 & 0 & \frac{\Delta_e(1-\frac{V_2k^2}{V_0})}{2}\\
      0 & \epsilon_{1/2}(k) & 0 & 0 & 0 & 0 & -\frac{\Delta_e(1-\frac{V_2k^2}{V_0})}{2} & 0\\
      0 & 0 & \epsilon_{1/2}(k) & 0 & 0 & \frac{\Delta_e(1-\frac{V_2k^2}{V_0})}{2} & 0 & 0\\
      0 & 0 & 0 & \epsilon_{3/2}(k) & -\frac{\Delta_e(1-\frac{V_2k^2}{V_0})}{2} & 0 & 0 & 0\\
      0 & 0 & 0 & -\frac{\Delta_e(1-\frac{V_2k^2}{V_0})}{2} & -\epsilon_{3/2}(k) & 0 & 0 & 0\\
      0 & 0 &\frac{\Delta_e(1-\frac{V_2k^2}{V_0})}{2} & 0 & 0 & -\epsilon_{1/2}(k) & 0 & 0\\
      0 & -\frac{\Delta_e(1-\frac{V_2k^2}{V_0})}{2} & 0 & 0 & 0 & 0 & -\epsilon_{1/2}(k) & 0\\
      \frac{\Delta_e(1-\frac{V_2k^2}{V_0})}{2} & 0 & 0 & 0 & 0 & 0 & 0 & -\epsilon_{3/2}(k)
\end{matrix}\right)\psi(\bk).
\eea

The mean-field transition temperatures are determined by the following gap equations:
\bea\label{eq:Tc}
\frac{1}{-V_0}=\sum_{n=\frac{1}{2},\frac{3}{2}}\int_0^{\Lambda}\frac{k^2dk}{2\pi^2}\tanh\frac{|\epsilon_{n}(k)|}{2T_e}\frac{\partial^2E^e_{n}(k)}{\partial{\Delta_e}^2}|_{\Delta_e=0},\\
\nonumber
\frac{1}{3V_2/2}=\sum_{n=\frac{1}{2},\frac{3}{2}}\int_0^{\Lambda}\frac{k^2dk}{2\pi^2}\tanh\frac{|\epsilon_{n}(k)|}{2T_o}\frac{\partial^2E^o_{n}(k)}{\partial{\Delta_o}^2}|_{\Delta_o=0},
\eea
where
\bea\label{eq:BdG_dispersion}
E^e_{n}(k)&=&\sqrt{\epsilon_n^2(k)+\Delta^2_e(1-\frac{V_2}{V_0}k^2)^2/{4}},\\
\nonumber
E^o_{1/2}(k)&=&\sqrt{\epsilon_{1/2}^2(k)+\frac{\Delta_o^2}{45}k^2},\\
\nonumber
E^o_{3/2}(k)&=&\sqrt{\epsilon_{1/2}^2(k)+\frac{\Delta_o^2}{5}k^2}
\eea
are the quasiparticle dispersions of $\hat{H}_{BdG}$.

We now provide a simple proof that as long as $V_0<0$, the system always energetically favors the conventional even parity pairing. The free energy of the two phases is given by
\bea
F_{o}(\Delta_o,T)&=&-\frac{2T}{(2\pi)^3}\sum_{n=1/2,3/2}\int{dk^3}\ln\cosh[E^{o}_n(k)/2T]+\frac{2|\Delta_o|^2}{3V_2},\\
\nonumber
F_{e}(\Delta_e,T)&=&-\frac{2T}{(2\pi)^3}\sum_{n=1/2,3/2}\int{dk^3}\ln\cosh[E^{e}_n(k)/2T]+\frac{|\Delta_e|^2}{-V_0}.
\eea
Assume $V_0<0$ and define $\Delta_1=\Delta_e/\sqrt{-V_0}$ and $\Delta_2=\sqrt{2}\Delta_o/\sqrt{3V_2}$ to make the two terms $2\Delta_o^2/(3V_2)$ and $\Delta_e^2/(-V_0)$ have the same functional dependence. Then we have
\bea
E_n^e(k,\Delta_1)&=&\sqrt{\epsilon_n^2(k)+\Delta_1^2|V_0|(1-\frac{V_2}{V_0}k^2)/4}\\
\nonumber
&=&\sqrt{\epsilon_n^2(k)+\Delta_1^2(|V_0|+\frac{V_2k^2}{2}+\frac{V_2^2k^4}{4|V_0|})},\\
\nonumber
E_{1/2}^e(k,\Delta_2)&=&\sqrt{\epsilon_{1/2}^2+\frac{V_2\Delta_2^2}{30}k^2},\\
\nonumber
E_{3/2}^e(k,\Delta_2)&=&\sqrt{\epsilon_{3/2}^2+\frac{3V_2\Delta_2^2}{10}k^2}.
\eea
Obviously we have
\bea
E_n^e(k,\Delta)>E_n^o(k,\Delta)
\eea
for $n=1/2,3/2$. This means that
\bea
F_o(\Delta,T)>F_e(\Delta,T)
\eea
for any $\Delta$ and $T$. Therefore the transition corresponding to $\Delta_e$ must happen at a higher temperature.
\end{appendix}
\end{document}